\documentclass{article}
\newcommand{\bfr}{\begin{flushright}}
\newcommand{\efr}{\end{flushright}}
 
\begin{document}
% \eqsec  % uncomment this line to get equations numbered by (sec.num)
\title{U($\infty$) Gauge Theory from Higher Dimensions
%\thanks{Presented at ...}%
% you can use '\\' to break lines
}
\author{Kiyoshi Shiraishi\\
%\address{
Institute  for  Cosmic Ray Research,  University of  Tokyo,\\
2-1  Midori-cho 3,  Tanashi-shi,  Tokyo  188,  Japan%}
}
\date{Int. J. Mod. Phys. {\bf A7} (1992)
pp. 6025--6037 }
\maketitle
\begin{abstract}
We show that classical U($\infty$) gauge theories can be obtained from
the dimensional reduction of a certain class of higher-derivative
theories. In general, the exact symmetry is attained in the limit of
degenerate metric; otherwise, the infinite-dimensional symmetry can be
taken as spontaneously broken. Monopole solutions are examined in the
model for scalar and gauge fields. An extension to gravity is also
discussed.
\end{abstract}
%\PACS{}

%%%%%%%%%%%%%%%%%%%%%%%%%%%%%%%%%%%%%%%%%%%%%%%%%%%
\section{Introduction}
%%%%%%%%%%%%%%%%%%%%%%%%%%%%%%%%%%%%%%%%%%%%%%%%%%%
The idea of attributing the internal symmetry of fields to the
symmetry of some compact space has grown in the last several decades.
In recent years, many authors have studied the possibility of
identifying the unified symmetry with the isometry of the extra spaces,
which is motivated by Kaluza-Klein gravity \cite{1} and superstring
theory.\cite{2} The local gauge symmetry in Kaluza-Klein theory
originates from the symmetry of extra space.

The symmetry is only a part of an infinitely large symmetry of extra
dimensions. In Ref.~\cite{3}, Dolan and Duff advocated an interpretation
that the large spatial symmetry is broken down to the symmetry of
massless states by spontaneous compactification. It is natural that the
infinite dimensional (Kac-Moody) symmetry is recovered in the infinite
limit of compactification scale, or decompactification.

Here we show the simplest example. Suppose that the $I$-th dimension is
compactified to a circle with circumference $L$. Then the general
transformation of the off-diagonal component of metric is
\begin{equation}
\delta g_{\mu I}=\nabla_\mu\xi_I+\nabla_I\xi_\mu\,.
\label{eq1}
\end{equation}
If we expand the parameter $\xi_M$ $(M=\mu, I)$ as
\begin{equation}
\xi_M=\sum_m\xi^m_M\, e^{imy/L}\,,
\end{equation}
where $y$ denotes the $I$-th coordinate, then the ``zero mode'' of
(\ref{eq1}) reduces to the gauge transformation of $A^0_\mu\equiv
g^0_{\mu I}$, i.e.
\begin{equation}
\delta A_{\mu}^0=\nabla_\mu\xi_I^0\,,
\end{equation}
while the transformation of the other field components are
\begin{equation}
\delta A_{\mu}^m=\nabla_\mu\xi_I^m+\frac{im}{L}\xi_\mu^m\,.
\label{1.4}
\end{equation}
Therefore one can say that the ``gauge'' transformation is broken in the
case of $m\ne 0$. The mass of the Kaluza-Klein excited state has the
same origin as the right-most term of (\ref{1.4}). The appearance of
mass indicates symmetry breaking! The infinite-dimensional symmetry is
restored in the decompactification limit, $L\rightarrow\infty$. For
details, such as the algebraic structure of the field transformation,
see Ref.~\cite{3}.%
\footnote{For the symmeuy breaking in higher dimensional Yang-Mills
theory, see also Ref.~\cite{4}.}

The study of the possible structure of underlying symmetry is very
useful in revealing the deeper mathematical importance of the model and
to give a clear insight into model-building.

Recently Floratos et al. \cite{5} offered a model with the gauge
symmetry of an infinite dimensional group. The symmetry, SU($\infty$)
is being eagerly investigated in the study of membrane theory.\cite{6}

A naive expectation suggests that the SU($\infty$) model may be derived
from some higher-dimensional model (at least in a certain limit),
because the model contains an infinite number of particle states.

In this paper we take a heuristic approach. Note that the Lagrangian
of the model is not necessarily of the same form as the dimensionally
reduced one. For instance, the Lagrangian which leads to
four-dimensional Yang-Mills theory does not have to be a Yang-Mills
Lagrangian in higher dimensions.

The organization of this paper is as follows. In Sec.~2, we briefly
review the SU($\infty$) Yang-Mills theory in order to make the present
paper self-contained. In Sec.~3, we will illustrate a simple model
which reduces to scalar field theory coupled to SU($\infty$) Yang-Mills
fields. As we will see there, the SU($\infty$) Yang-Mills field comes
from a U(1) gauge field in higher dimensions. We will show that the
two-dimensional SU($\infty$) Yang-Mills theory is obtained from the
four-dimensional U(1) model. In our models, the abelian gauge field
remains as a zero mode after the dimensional reduction. Thus the
symmetry group is actually SU($\infty$)$\times$U(1) or U($\infty$). In
general dimensions, a $D$-dimensional U($\infty$) theory is derived from
a ($D+2$)-dimensional higher-derivative theory in the
degenerate-metric limit. In Sec.~5, the Bogomol'nyi-Prasad-Sommerfield
monopole \cite{7} is constructed from the scalar and gauge theory
introduced in Secs.~3 and 4. An SU(2) subgroup is utilized in the
construction. Section 6 is devoted to summary and outlook, including
discussion of extension to a fermionic model and gravity.

%%%%%%%%%%%%%%%%%%%%%%%%%%%%%%%%%%%%%%%%%%%%%%%%%%%
\section{Review of SU($\infty$) Yang-Mills Theory}
%%%%%%%%%%%%%%%%%%%%%%%%%%%%%%%%%%%%%%%%%%%%%%%%%%%
It is well known that SU($\infty$) symmetry may be considered as a
volume-preserving diffeomorphism of some two-dimensional surface, which
can be identified with a membrane.\cite{6}

The SU($\infty$) Yang-Mills theory \cite{5} has been put forward in the
context of the study of membrane theory.\cite{6} To make our discussion
self-contained, we review the SU($\infty$) Yang-Mills theory \cite{5} in
this section.

We consider U(1) gauge fields which have dependence on ``extra
coordinates'' $\theta$ and $\varphi$ as well as the space-time
coordinates. $\theta$ and $\varphi$ span the standard coordinate system
of the two-sphere. The gauge fields can be expanded in the following
form:
\begin{equation}
A_\mu(x^\nu;\theta,\varphi)=\sum_{l=1}^\infty\sum_{m=-l}^l
A_\mu^{lm}(x^\nu)Y_{lm}(\theta,\varphi)\,,
\label{2.1}
\end{equation}
where $Y_{lm}(\theta,\varphi)$ is the spherical harmonic function.

SU($\infty$) field strength and gauge transformations are defined by
\begin{equation}
\tilde{F}_{\mu\nu}=\partial_\mu A_\nu-\partial_\nu A_\mu+
\{A_\mu, A_\nu\}\,,
\end{equation}
\begin{equation}
\delta A_\mu(x^\nu;\theta,\varphi)=\partial_\mu
\omega(x^\nu;\theta,\varphi)+
\{A_\mu, \omega\}\,,
\end{equation}
\begin{equation}
\delta \tilde{F}_{\mu\nu}=\{\tilde{F}_{\mu\nu}, \omega\}\,,
\end{equation}
where
\begin{equation}
\{f, g\}=\frac{\partial f}{\partial\cos\theta}\frac{\partial
g}{\partial\varphi}-\frac{\partial
f}{\partial\varphi}\frac{\partial g}{\partial\cos\theta}=
-\frac{1}{\sqrt{g^{(2)}}}\varepsilon^{mn}\partial_mf\partial_ng\,.
\label{2.5}
\end{equation}
Here $g^{(2)}$ is the determinant of the standard metric of $S^2$.

The bracket corresponds to the commutator in the usual Yang-Mills
theory in the matrix-valued representation, while integration with
respect to the extra coordinates corresponds to taking trace.

The algebra formed by the brackets can be identified with SU($\infty$)
symmetry. This symmetry arises from the diffeomorphisms of $S^2$. The
diffeomorphisms of $T^2$, the two-torus, can also lead to an
SU($\infty$) algebra, although the process of taking the limit
$N\rightarrow\infty$ in SU($N$) is slightly different from the case of
$S^2$.\cite{8}

It is natural to suppose that such a theory is derived from some
nonlinear theory in higher dimensions where $\theta$ and $\varphi$ are
the coordinates of extra spaces. In the next section, we will describe
a model which leads to a scalar field coupled to the SU($\infty$)
Yang-Mills field after the dimensional reduction as the simplest
example.

%%%%%%%%%%%%%%%%%%%%%%%%%%%%%%%%%%%%%%%%%%%%%%%%%%%
\section{Dimensional Reduction of a Scalar Model}
%%%%%%%%%%%%%%%%%%%%%%%%%%%%%%%%%%%%%%%%%%%%%%%%%%%
In this section, we concentrate our attention on the derivation of
classical scalar field theory with local SU($\infty$) symmetry from
dimensional reduction. 

The key step is to find the antisymmetric tensor
in (\ref{2.5}). We wish to consider that
the tensor arises ``spontaneously'' in  higher dimensions.  To this 
end,  we  utilize  the vacuum  expectation  value  of  the  U(1) 
gauge  field  strength.

First,  we  consider  the  Lagrangian
\begin{equation}
L_S=\delta_{DEF}^{ABC}F^{DE}F_{AB}\partial^F\phi\partial_C\phi\,,
\label{3.1}
\end{equation}
where the totally antisymmetric Kronecker's symbol is defined by
\begin{equation}
\delta_{DEF}^{ABC}=3!\delta^A_{[D}\delta^B_E\delta^C_{F]}\,,
\end{equation}
and  the  generalization  with more  suffixes  is  straightforward. 
Our  simple  model  for  a scalar  field is  based  on  the  Lagrangian
(\ref{3.1}).

The Lagrangian is rewritten  as
\begin{equation}
L_S=2(F^{AB}F_{AB}\partial^C\phi\partial_C\phi-
2F^{AC}F_{AB}\partial^B\phi\partial_C\phi)\,.
\end{equation}
Thus we can immediately see that the kinetic term of the scalar is
generated if the field strength acquires an expectation value.

We consider the Lagrangian in ($D+2$)-dimensional space-time. We use
indices $m, n, \dots$ for compact two-dimensional space, say, $S^2$ or 
$T^2$, while we use $\mu, \nu, \dots$ for the $D$-dimensional
space-time. For the present, we take a unit scale for the size of the
compact space.

Here we assume a vacuum expectation value for the extra components of
the field strength
\begin{equation}
\langle F_{mn}\rangle=\frac{1}{q}\varepsilon_{mn}\,,
\label{3.4}
\end{equation}
where $q$ is a constant.  Note that  $1/q$ is often quantized as an
integer or half-integer (in the  unit  of the  compactification 
volume)  for  topological  reasons.\cite{10}

Further we set $\partial_\mu A_m=0$. If  we leave $A_m$'s , these will
induce residual, nonminimal interacting  scalars  other  than 
SU($\infty$)  gauge fields, just  as  in  Kaluza-Klein  theory.

Classifying  the  suffices  into  $\mu$'s and $m$'s,  in  the 
Lagrangian (\ref{3.1}),  we find
\begin{eqnarray}
\frac{1}{2}\delta^{ABC}_{DEF}F^{DE}F_{AB}\partial^F\phi\partial_C\phi&=&
(F^{mn}F_{mn})\partial^\mu\phi\partial_\mu\phi-4
F^{mn}F_{m\nu}\partial^\nu\phi\partial_n\phi\nonumber \\
&+&2(F^{m\mu}F_{m\mu}\partial^n\phi\partial_n\phi-
F^{\mu m}F_{\mu n}\partial^n\phi\partial_m\phi)\nonumber \\
&+&(F^{\mu\nu}F_{\mu\nu})\partial^m\phi\partial_m\phi-4
F^{\mu m}F_{\mu\nu}\partial^\nu\phi\partial_m\phi\nonumber \\
&+&2(F^{m\mu}F_{m\mu}\partial^\nu\phi\partial_\nu\phi-
F^{m\mu}F_{m\nu}\partial^\nu\phi\partial_\mu\phi)\nonumber \\
&+&(F^{\mu\nu}F_{\mu\nu})\partial^\lambda\phi\partial_\lambda\phi-2
F^{\mu\lambda}F_{\mu\nu}\partial^\nu\phi\partial_\lambda\phi\,.
\label{3.5}
\end{eqnarray}

We first pay attention to the first line on the right-hand side of
Eq.~(\ref{3.5}). Substitution of the ans\"atze,
$F_{mn}=q^{-1}\varepsilon_{mn}$ and $\partial_\mu A_m=0$ into this
part yields
\begin{eqnarray}
& &(F^{mn}F_{mn})\partial^\mu\phi\partial_\mu\phi-4
F^{mn}F_{m\nu}\partial^\nu\phi\partial_n\phi\nonumber \\
& &+2(F^{m\mu}F_{m\mu}\partial^n\phi\partial_n\phi-
F^{\mu m}F_{\mu n}\partial^n\phi\partial_m\phi)\nonumber \\
&
&\Rightarrow\frac{2}{q^2}(\partial_\mu\phi-q\varepsilon^{mn}\partial_m
A_\mu\partial_n\phi)^2\,.
\label{3.6}
\end{eqnarray}
This term is just the kinetic term of the scalar boson in
the ``adjoint'' representation of SU($\infty$) symmetry,\cite{9}
$(D_\mu\phi)^2=(\partial_\mu\phi+q\{A_\mu, \phi\})^2$. Note that a
``neutral'' scalar field is added, which comes from the zero mode of the
expansion on the extra coordinates.

In general cases, the remaining terms in (\ref{3.5}) do not have the
SU($\infty$) local symmetry. Nevertheless, all the scalar modes are
massless in the expansion on the extra coordinates, unlike ordinary
Kaluza-Klein theory; this is because there is no
$(\partial_m\phi\partial^m\phi)$ term in the Lagrangian.

 Comparing the terms in the right'hand side of (\ref{3.5}) with each
other, we find a difference in the number of the extra-space indices
contained in each term. The terms in the first line of the right-hand
side contain two couples of repeating indices of $m, n$, while the
second line contains one couple and the third none.

If we denote the radius of the extra space as $b$, instead of the unit
scale, the metric has a dependence $g_{mn}\approx b^2$ and then 
$g^{mn}\approx b^{-2}$. Since the contraction of the extra-space
indices is performed by use of $g^{mn}$, the first line, i.e.
(\ref{3.6}) becomes the dominant contribution for small $b$. At the same
time, of course, we must rescale $q$ so as to keep $q/b^2$ finite.

Therefore in the limit of the degenerate metric $b\rightarrow 0$ with an
appropriate normalization of fields, we obtain an exact SU($\infty$)
gauge symmetric model. Our model forms a good contrast to gauge theory
from ordinary Kaluza-Klein gravity.\cite{3} The latter exhibits large
symmetry in the decompactification limit, while ours shows symmetric
form in the degenerate compactification limit.

Our model has unbroken SU($\infty$) local symmetry in a particular
dimension, even if the compactiflcation scale is finite. This can
easily be seen in three dimensional space-time; in this case the
Lagrangian of our model can be written as
\begin{equation}
L_S\approx\left(\frac{1}{\sqrt{|g|}}\varepsilon^{ABC}F_{AB}\partial_C\phi\right)^2\,.
\end{equation}
This gives only the contribution (\ref{3.6}) when the dimensional
reduction with the nonzero field strength is carried out.

In the next section, we will construct a model Lagrangian which
induces the kinetic term for an SU($\infty$) Yang-Mills field.

%%%%%%%%%%%%%%%%%%%%%%%%%%%%%%%%%%%%%%%%%%%%%%%%%%%
\section{U($\infty$) Yang-Mills Theory from Dimensional Reduction}
%%%%%%%%%%%%%%%%%%%%%%%%%%%%%%%%%%%%%%%%%%%%%%%%%%%
As we have seen in the previous section, we can construct a model with
local SU($\infty$) symmetry from dimensional reduction. In this section,
we will write down the higher-dimensional Lagrangian which produces the
SU($\infty$) Yang-Mills kinetic term by reduction.

We note that the models we are studying contain higher derivatives;
moreover, the kinetic terms in $D$ dimensions appear only if the field
strengths take nonzero values. In this sense, our models can be
compared to a gauge-theory version of a ``pregeometric'' theory of
gravity.

Another point to notice is that the action of our model can be written
in terms of differential forms.\cite{11} For example, the action of the
scalar model in the previous section is expressed as
\begin{equation}
\int(F\wedge d\phi)\wedge{}^*(F\wedge d\phi)\,,
\end{equation}
where  $F$  is  the  U(1)  curvature  two-form and  ${}^*$  denotes 
the  dual.

Now,  the  model  which  leads  to  Yang-Mills  must have  a  highly 
symmetric  style. We  can first  generalize  the  Lagrangian
(\ref{3.1})  very  naturally.  Then  we  obtain
\begin{equation}
L_{YM}=\delta^{ABCD}_{EFGH}F^{EF}F_{AB}F^{GH}F_{CD}\,,
\end{equation}
and this is proportional to
\begin{equation}
(F_{AB}F^{AB})^2 - 2F^A{}_BF^B{}_CF^C{}_DF^D{}_A\,.
\end{equation}
This form is also motivated by the Euler form in which
$R_{ABCD}=F_{AB}F_{CD}$ is substituted. It is easy to rewrite the action
in terms of differential forms as
\begin{equation}
\int(F\wedge F)\wedge{}^*(F\wedge F)\,,
\label{4.4}
\end{equation}
which seems to be an extension of the Maxwell action
\begin{equation}
\int F\wedge{}^*F\,.
\end{equation}

In four dimensions, the Lagrangian (\ref{4.4}) can be rewritten as
\begin{equation}
L_{YM}\approx\left(\frac{1}{\sqrt{|g|}}\varepsilon^{ABCD}F_{AB}
F_{CD}\right)^2\,.
\label{4.6}
\end{equation}
Note that this is not a ``topological'' Lagrangian even in four
dimensions.

Here we take the same ansatz (\ref{3.4}) for the field strength. This
expectation value is consistent with the equation of motion in the
vacuum. This is because the suffices are combined by Kronecker's delta
in the action and the excessive overlapping of the
suffixes  is  avoided.

For  the  same  reason,  mass  terms  for the  gauge  fields are  absent  after  dimensional
reduction,  unlike  ordinary  Kaluza-Klein  type  theories.

We  first  analyze  the  model  in  $2+2$  dimensions,  i.e.   we 
adopt  the  Lagrangian (\ref{4.6}).  Assuming  $\partial_\mu A_m=0$
($\mu=0, 1$),  we  obtain
\begin{eqnarray}
L_{YM}(D=2)&\approx&\frac{1}{q^2}(\partial_0A_1-\partial_1A_0
+q\{A_0,A_1\})^2\nonumber
\\ &\equiv&\frac{1}{2}\frac{1}{q^2}\tilde{F}_{01}^2\,,
\label{4.7}
\end{eqnarray}
for compactification with a unit scale. Here we get the Lagrangian of
two-dimensional classical U($\infty$) Yang-Mills theory.

The reason why we have called the symmetry group U($\infty$) rather than
SU($\infty$ ) is as follows. For the SU($\infty$) Yang-Mills theory,
the expansion by harmonics begins with $l=1$ [see (\ref{2.1})]. In other
words, we omit the zero mode of the two-manifold. In the derivation from
higher dimensions, however, a U(1) is included by the zero mode.
Therefore, the gauge symmetry obtained after the reduction is
SU($\infty$)$\times$U(1) or U($\infty$).

Next, in general dimensions, we get the U($\infty$) gauge theory from
dimensional reduction in the degenerate limit, the scale of the extra
space $b\rightarrow 0$, as in the scalar model in the previous section.

The classification of the Lagrangian done as in the previous section
is as follows:
\begin{eqnarray}
& &(F_{AB}F^{AB})^2 - 2F^A{}_BF^B{}_CF^C{}_DF^D{}_A=
(F_{mn}F^{mn})^2-2
F^m{}_nF^n{}_pF^p{}_qF^q{}_m\nonumber \\
& &\qquad\qquad\qquad\qquad\qquad+4(F_{mn}F^{mn})(F_{p\mu}F^{p\mu})-8
F^\mu{}_mF^m{}_nF^n{}_pF^p{}_\mu\nonumber \\
&
&\qquad\qquad\qquad\qquad\qquad+2(F_{mn}F^{mn})(F_{\mu\nu}F^{\mu\nu})-8
F^\mu{}_\nu F^\nu{}_mF^m{}_nF^n{}_\mu\nonumber \\ &
&\qquad\qquad\qquad\qquad\qquad+4[(F^{m\mu}F_{m\mu})^2- F^\mu{}_m
F^m{}_\nu F^\nu{}_nF^n{}_\mu]\nonumber \\ &
&\qquad\qquad\qquad\qquad\qquad+4(F_{m\mu}F^{m\mu})(F_{\nu\lambda}F^{\nu\lambda})-8
F^\mu{}_\nu F^\nu{}_\lambda F^\lambda{}_mF^m{}_\mu\nonumber \\ &
&\qquad\qquad\qquad\qquad\qquad+(F_{\mu\nu}F^{\mu\nu})^2-2 F^\mu{}_\nu
F^\nu{}_\lambda F^\lambda{}_\sigma F^\sigma{}_\mu\,.
\label{4.8}
\end{eqnarray}

The first and second lines on the right-hand side of Eq.~(\ref{4.8})
vanish regardless of the dimension $D$, when the ans\"atze for the extra
gauge fields are assumed. The third line becomes the Yang-Mills term
$(4/q^2)\tilde{F}_{\mu\nu}^2$ when the ans\"atze are taken into
consideration. The fourth and fifth lines remain if the
compactification scale $b\ne 0$ and the dimension $D+2\ne 4$.

Here we consider generalization to more higher-derivative terms. Let
us generalize the form (\ref{4.6}) to a term including more
derivatives, i.e. consider
\begin{equation}
(\varepsilon^{ABCDEF\cdots}F_{AB}F_{CD}F_{EF}\cdots)^2\,.
\label{4.9}
\end{equation}
This generic structure appears in the expansion of the determinant of
some set of matrices including $F_{AB}$. These terms remind us of the
Born-Infeld action,\cite{12}%
\footnote{For further references, see Ref.~\cite{13}.}
\begin{equation}
S_{BI}=\int
d^{D+2}x\,\frac{C}{\alpha^2}[\sqrt{-\det(\bar{g}_{AB}+\alpha F_{AB})} 
-\sqrt{-\bar{g}}]\,,
\end{equation}
where $\alpha$ is a coupling which has the dimensions of
(mass)${}^{-2}$, while the constant $C$ has the dimensions of
(mass)${}^{D-2}$. This action can be expanded into terms like
(\ref{4.9}), with respect to small $\alpha$.

Further we take the background geometry of partially compactified
space as usual:
\begin{equation}
\bar{g}_{AB}=\left(\begin{array}{cc}
g_{\mu\nu} & 0\\
0 & b^2\tilde{g}_{mn}
\end{array}\right)\,,
\end{equation}
where $b$ is the radius of compact space and $\mu, \nu$ run over $0,
1,\dots, D$ while $m, n$ denote the extra-coordinate indices as usual.
$\tilde{g}_{mn}$ is the metric on the extra space with unit scale. If we
take the field strength $F_{mn}=q^{-1}\varepsilon_{mn}$ as the
background and set $\partial_\mu A_m=0$ by hand as previously, we find
the reduced Lagrangian in the limit of $b\rightarrow 0$ is proportional
to
\begin{equation}
\frac{C}{\alpha^2}\left[\sqrt{-\frac{\alpha^2}{e^2}
\det(g_{\mu\nu}+\alpha
\tilde{F}_{\mu\nu})}\right]\,,
\end{equation}
where
\begin{equation}
\tilde{F}_{\mu\nu}\equiv\partial_\mu A_\nu-\partial_\nu
A_\mu+e\{A_\mu, A_\nu\}\,,
\end{equation}
and $e=q/b^2$.

Now up to overall normalization and cosmological constant to be
adjusted, we get the U($\infty$) Born-Infeld Lagrangian with the
coupling constant $e$. 

It is notable that if we first take the limit $\alpha\rightarrow
0$, we obtain only a usual Kaluza-Klein reduction of U(1) fields.

In this section, we have considered the pure Yang-Mills sector of
U($\infty$) theory. We will consider classical solutions in a
U($\infty$) Yang-Mills-Higgs system in the following section.

%%%%%%%%%%%%%%%%%%%%%%%%%%%%%%%%%%%%%%%%%%%%%%%%%%%
\section{BPS Monopole in the U($\infty$) Theory}
%%%%%%%%%%%%%%%%%%%%%%%%%%%%%%%%%%%%%%%%%%%%%%%%%%%
Topological objects provide various views of field theory, both
classical and quantum, and sometimes of the nonperturbative nature of
it.

In the present section, we will discuss classical solutions in a
U($\infty$) Yang-Mills-Higgs system.

In Sec. 3, we introduced a U($\infty$) scalar theory. Using this scalar
as a Higgs field, we can consider a Yang-Mills-Higgs system. Although
we can construct a potential term for the scalar field from dimensional
reduction, we do so proceed in this paper. 

The Lagrangian we consider is
\begin{eqnarray}
L_{YMH}&=&\frac{\beta}{4}(F^{AB}F_{AB}\partial^C\phi\partial_C\phi-
2F^{AC}F_{AB}\partial^B\phi\partial_C\phi)\nonumber \\
&+&\frac{\beta}{16}[(F_{AB}F^{AB})^2 - 2F^A{}_BF^B{}_CF^C{}_DF^D{}_A]\,,
\label{5.1}
\end{eqnarray}
where $\beta$ is a coupling constant which has dimension
(mass)${}^{-2}$.

We assume six-dimensional space-time. We must take care of the
contribution of the nonminimal residual interaction which appears from
the dimensional reduction.

We examine a static monopole configuration in the system. For
simplicity, we investigate the Prasad-Sommerfield limit,\cite{7} i.e.
the limit of no Higgs-self-coupling.
Thus we need not worry about the potential term.

To construct monopole configurations, we pick out an SU(2) subgroup in the
U($\infty$) group. This can easily be found; the $l=1$ spherical
harmonics $Y_{lm}(\theta, \varphi)$ $(m=-1, 0, 1)$ play the role of
generators of SU(2), in the case of the algebra formed by the bracket
(\ref{2.5}). Thus we assume that the gauge fields and scalar fields take
nonzero values in this subgroup sector, i.e.
\begin{equation}
\phi=\sum_{a=1}^3\phi^a(x^i)T^a(\theta,\varphi)\,,\quad
A_i=\sum_{a=1}^3 A_i^a(x^j)T^a(\theta,\varphi)\,,
\label{5.2}
\end{equation}
where $T^a$  $(a=1, 2, 3)$ are defined as
\begin{equation}
T^1=\frac{1}{\sqrt{2}}(Y_{1\,1}+iY_{1\,-1})\,,\quad
T^2=\frac{1}{\sqrt{2}}(Y_{1\,1}-iY_{1\,-1})\,,\quad
T^3=Y_{1\,0}\,.
\end{equation}
In (\ref{5.2}), $x^i$ $[i( j) = 1, 2, 3]$ denotes the spatial
coordinates. Besides these fields, of course, the ``monopole''
configuration of the gauge field \cite{10} exists in the extra space, as
Eq.~(\ref{3.4}).

Now the bracket is defined in terms of the extra coordinates with unit
radius of the sphere. Then the general compactification scale $b$ is
absorbed into the coupling, $e = q/b^2 =$ a finite constant. If we
require the condition of no physical singularity in the background
gauge field which generates $F_{mn}$, $e$ must be quantized as $2/n$
($n:$ integer).\cite{10}

We make the following ``spherical ansatz'' for the solution in the
explicit form:
\begin{equation}
\phi^a=\frac{x^a}{er^2}H(evr)\,,\quad
A_i^a=-\varepsilon_{aij}\frac{x^j}{er^2}[1 - K(evr)]\,,
\end{equation}
where $r=|x|$ and $a,  i,  j  =  1,  2,  3$. $v$ is an expectation
value for $|\phi|$ which is taken at spatial  infinity.  In  our  case, 
$\phi$ gives  a  one-to-one  mapping  from  a  point  of  spatial
infinity  ($S^2$)  to  a  point  on  the  extra  sphere  $S^2$. 

$H$ and $K$ must  be  subject  to  the boundary  condition
\begin{equation}
H(\xi)\rightarrow\xi\,,\quad K(\xi)\rightarrow 0\quad
\mbox{when~} r\rightarrow\infty\,,
\end{equation}
where  $\xi =  evr$.

For the  time-independent   ansatz,   the   energy   of   the  
system   is   obtained  after integrating  over the  extra
coordinates,  $\theta$ and $\varphi$,  as  follows:
\begin{eqnarray}
& &\!\!\!\!\!\!\!\!\!\!\!\!
 E=\frac{8\pi^2 v\beta}{e^3b^2}\int_0^\infty \!\!\!\!
d\xi\left\{
\left(\frac{dK}{d\xi}\right)^2+\frac{1}{2}
\left(\frac{dH}{d\xi}-\frac{H}{\xi}\right)^2+
\frac{(K^2-1)^2}{2\xi^2}+\frac{K^2H^2}{\xi^2}\right.\nonumber \\
& &\!\!\!\!\!\!\!\!\!\!
+\frac{(evb)^4}{30}\left[15H^2
\left(\frac{dK}{d\xi}\right)^2+20\frac{1}{2}
\left(\frac{dH}{d\xi}-\frac{H}{\xi}\right)^2(1-K)^2+18H^2(1-K)^2\right.
\nonumber \\
& &+\left.\left.14\left(\frac{dK}{d\xi}\right)^2(1-K)^2
+64\xi^2(1-K)^4\right]\right\}\,.
\label{5.5}
\end{eqnarray}

At a glance, we can see the energy is always positive; moreover we
find that the energy for finite $b$ is larger than the mass of the
Prasad-Sommerfield monopole, $E_0= 8\pi^2v\beta/(e^3b^2)$ in the present
case.

The field equation is obtained by the variational principle. If we
consider the contribution of the second line of (\ref{5.5}) as a
perturbation. we expand the solution as $K = K_0+ \epsilon K_1+\cdots$
and $H = H_0+\epsilon H_1+\cdots$, where $\epsilon= (evb)^4/30$. $K_0$
and $H_0$ are the solutions for the equation where we set $\epsilon= 0$,
which satisfy the Bogomol'nyi equations.\cite{7} We find the asymptotic
behavior of $K_1$ and $H_1$ at spatial infinity and in the vicinity of
the origin as follows:
\begin{eqnarray}
& &K_1\approx
e^{-\xi}\,,\quad\frac{H_1}{\xi}\approx\frac{1}{\xi}\quad\mbox{at spatial
infinity}\,,\nonumber
\\ & &K_1\approx \xi^2\,,\quad H_1\approx\xi^2\quad\mbox{near the
origin}\,.
\end{eqnarray}

Unfortunately, the exact solution can be obtained only by numerical
calculation. But here it is sufficient to know that the correction to
the energy due to the residual interaction is of the order of
$\epsilon E_0$.

To summarize, we have found the monopole configuration in the system
which is described by the higher-derivative action (\ref{5.1}) in six
dimensions. The solution is a nontrivial configuration of the gauge and
scalar fields which belong to the SU(2) subgroup of the U($\infty$)
gauge group which becomes an exact symmetry if $b = 0$. We find the
energy of the monopole suffers the correction due to the residual
Wteraefions if $b \ne 0$. The order of the correction is
$\approx\epsilon E_0$, where $\epsilon= (evb)^4/30$.

Apparently, the lowest energy solution is attained if $b=0$ for fixed
$\beta/b^2$. So we again see that $b=0$ is a special point in our model.

As another example of a classical solution, we can consider locally
concentrated magnetic flux using a Nielsen-Olesen vortex in $D+2$
dimensions. We can construct the $D$-dimensional ``string'' \cite{14}
in which U($\infty$) Yang-Mills gauge bosons exist. The properties of
this solution will be reported in a separate publication.

%%%%%%%%%%%%%%%%%%%%%%%%%%%%%%%%%%%%%%%%%%%%%%%%%%%
\section{Summary and Outlook}
%%%%%%%%%%%%%%%%%%%%%%%%%%%%%%%%%%%%%%%%%%%%%%%%%%%
To summarize : we have obtained the $D$-dimensional U($\infty$)
Yang-Mills action from dimensional reduction of a ($D+2$)-dimensional
theory. The partially ``degenerate'' metric tensor induces the large
symmetry. This is compared with topological field theory,\cite{15} in
which the metric can be taken as completely degenerate. There is no
massive excitation mode \`a la Kaluza-Klein regardless of the size of
the extra space.
The inverse of the coupling constant is quantized if we require regular
background gauge fields in many cases.\cite{10}

Our model Lagrangian is very similar to the term which appears in the
expansion of the Born-Infeld Lagrangian.\cite{12} If the Born-Infeld
action is derived from (open) string theory,\cite{16} $\alpha$ is
related to the slope parameter. Therefore we can speculate that the
large symmetry may have a relation to both the compactification scale
and the ``stringy'' scale.

A monopole configuration has been considered. The lowest energy solution is
realized when the residual interactions vanish, or $b =0$.

Now we consider extensions to other fields. A possible extension to a
fermion model which leads to coupling to the SU($\infty$) Yang-Mills
field is represented as
\begin{equation}
L_F=i\delta^{ABC}_{DEF}\bar{\psi}\Gamma^{DEF}F_{AB}D_C\psi
\approx i\bar{\psi}\Gamma^{ABC}F_{AB}D_C\psi\,,
\end{equation}
where $\Gamma^{ABC}$ is the antisymmetrized gamma matrix. This example
is the simplest one which yields a minimal coupling to the SU($\infty$)
Yang-Mills field after the reduction.

A naive extension to gravity is as follows:
\begin{eqnarray}
L_G&=&-\gamma\delta^{ABCD}_{EFGH}F^{EF}F_{AB}R^{GH}{}_{CD}\nonumber \\
&\approx&-\gamma (F^{CD}F_{AB} R^{AB}{}_{CD} - 4F^{CB} F_{AB} R^A_C +
F^{AB}F_{AB} R)\, , 
\label{6.2}
\end{eqnarray}
where $\gamma$ is a coupling constant.

Regrettably, the Lagrangian has only one massless ``graviton'' even if
the degenerate limit of the metric is taken. Thus the
infinite-dimensional symmetry is always broken.

Nevertheless, it is known that the classical solution of the model
described by the Lagrangian (\ref{6.2}) has remarkable properties. In
Ref.~\cite{17}, Yoshida and the present author showed that the
cosmological solutions derived from the Lagrangian are
obtained in analytical forms for many cases of six-dimensional
space-time. In the solutions, the inverse square of the radius of the
extra two dimensions behaves similarly to the scale of the extra circle
in five-dimensional Kaluza-Klein gravity. The correspondence holds in
the static spherical solutions.\cite{18} Whether this fact is connected
to some deep physical insight or not is as yet obscure.

The Lagrangian (\ref{6.2}) is very akin to the Euler form,
\begin{equation}
R^{CD}{}_{AB}R^{AB}{}_{CD}- 4R^C_AR^A_C + R^2\,.
\end{equation}
Indeed the dimensional reduction of the Euler form yields (\ref{6.2}) as
a part of it.\cite{19} It is known that six-dimensional Euler form
gravity leads to an infinite number of massless gravitons in four
dimensions if background space is partially compactified to $M_4\times
S^2$.\cite{20,21,22} Studying the quantum aspects of the model in four
dimensions is very attractive. We plan to analyze the Euler form
gravity \cite{20,21,22} and its quantum nature. A generic field theory
in higher dimensions suffers from disastrous UV divergence. If the role
of extra space is merely to generate a large symmetry, however, the
quantum nature of the theory is expected to be remarkably improved.
Both $1/D$ \cite{23} and $1/N$\cite{24} expansion techniques may be
available to study nonperturbative effects, where the number of
gravitons is regularized to be $N$. Further we wonder if only one
graviton remains massless and all the others become decoupled from
physical spectra. We hope our model of U($\infty$) Yang-Mills theory
may offer a toy model of such a higher-derivative theory.

The generalization to the case with higher-rank antisymmetric tensor
can be considered. New types of symmetry may be reduced from such
theories; it is an interesting possibility.

The models we have treated in this paper lead to classical U($\infty$)
models after dimensional reduction. This statement is at the same level
as in the case of five-dimensional Kaluza-Klein theory.\cite{25} In the
analyses of the previous sections we set $\partial_\mu A_m=0$, while in
Kaluza-Klein theory the pure Maxwell term appears if the radius of the
extra space is set to be constant. Thus when we consider various
aspects of our models we must be careful in treating the residual
interactions of $A_m$. The treatment of zero modes of Am will be
reported elsewhere.

\section*{Note Added}
After completion of this work, we became aware of the paper,\cite{26}
which is concerned with the SU($\infty$) gauge theory.

\section*{Acknowledgment}
The author is happy to thank Iwanami F\=ujukai for financial aid.

%%%%%%%%%%%%%%%%%%%%%%%%%%%%%%%%%%%%%%%%%%%%%% 

%%%%%%%%%%%%%%%%%%%%%%%%%%%%%%%%%%%%%%%%%%%%%
\end{document}